# An Extendible, Graph-Neural-Network-Based Approach for Accurate Force Field Development of Large Flexible Organic Molecules

Xufei Wang, Yuanda Xu,[a] Han Zheng,[c] Kuang Yu[c]*

a. Princeton University, Program of Applied and Computational Mathematics

c. Tsinghua-Berkeley Shenzhen Institute (TBSI), Institute of Materials Research (iMR),

Tsinghua Shenzhen International Graduate School (TSIGS), Tsinghua University.

ABSTRACT An accurate force field is the key to the success of all molecular mechanics simulations on organic polymers and biomolecules. Accuracy beyond density functional theory is often needed to describe the intermolecular interactions, while most correlated wavefunction (CW) methods are prohibitively expensive for large molecules. Therefore, it posts a great challenge to develop an extendible ab initio force field for large flexible organic molecules at CW level of accuracy. In this work, we face this challenge by combining the physics-driven nonbonding potential with a data-driven subgraph neural network bonding model (named sGNN). Tests on polyethylene glycol polymer chains show that our strategy is highly accurate and robust for molecules of different sizes. Therefore, we can develop the force field from small molecular fragments (with sizes easily accessible to CW methods) and safely transfer it to large polymers, thus opening a new path to the next-generation organic force fields.

I. INTRODUCTION

Undoubtedly, molecular force field is the key component underlying most molecular dynamics (MD) and Monte Carlo (MC) simulations. A reliable organic molecular force field is of paramount importance for the studies of biomolecules, small molecule drugs, polymers, and covalent organic frameworks (COF) etc. The state-of-the-art organic force fields mostly used to date (i.e., OPLS-AA,[1] AMBER,[2] CHARMM[3] etc.) are based on very simple functional forms (*e.g.*, Coulomb's law in conjunction with Lennard-Jones potentials) with empirical parameters. While they are explicitly fitted to reproduce macroscopic properties, they do not represent the microscopic details of the potential energy surface (PES) faithfully. Consequently, the accuracy and the transferability of these empirical force fields are largely limited. One often needs to reparametrize the force field for different systems, which is not only cumbersome, but also detrimental to the predictive power of the model.

Another line of research that has been increasingly popular is to build force field purely based on *ab initio* data. Unfortunately, density functional theory (DFT) is often not accurate enough to describe the intermolecular interactions, so high-level correlated wavefunction (CW) methods such as Moller-Plesset perturbation theory (MP) or Coupled Cluster theory (CC) are needed. However, most CW methods scale poorly with system size, making the direct evaluation of the high-dimensional PES prohibitively expensive. It has been a great challenge to develop a true *ab-abinito*-based force field that is accurate for both small molecule clusters and condensed phases.

The real difficulty in this task is how to learn information from low-dimensional small calculations, and extrapolate it to high-dimensional large systems. Such extrapolation relies heavily on the deep understanding to all the relevant physics and is highly nontrivial.

In the past decade, we have made essential progresses towards a physics-driven intermolecular (nonbonding) potential for organic molecules. Through time-dependent DFT (TD-DFT) calculations, we can obtain the charge density susceptibility matrix of the molecule. Then utilizing multipole expansion and proper localization methods, we can compute the asymptotic atomic parameters including multipole moments, polarizabilities, and dispersion coefficients.[4–6] Then, the residual medium- and short-range interactions can be fit term-by-term to Symmetry-Adapted Perturbation Theory (SAPT) dimer calculations.[7–9] This procedure possesses several advantages: 1. The atomic parameters resulting from TD-DFT are very accurate in long range (comparable with CCSD(T)); 2. The atomic parameters are physically meaningful, thus highly transferrable across different physical and chemical environments; and 3. The computational cost is relatively low as only monomer and dimer calculations are involved. However, its application is still limited to small or rigid molecules so far, due to two difficulties: 1. The atomic parameters can be conformation-dependent; and 2. The corresponding intramolecular (bonding) terms with comparable accuracies are yet to be developed. While the first issue can be solved by utilizing fluctuating atomic parameters, we will tackle the second issue in this work.

In the conventional organic force fields, the bonding energies are expressed as a direct sum of a series of uncoupled internal coordinates (*i.e.*, bond lengths, angles, dihedrals, and impropers *etc.*). Typically, anharmonicities and the nonlocal couplings between different internal coordinates are neglected, which significantly limits its accuracy. It has been shown that such couplings between different dihedrals play an important role in determining the conformations of large molecules.[10]

Meanwhile, the separation between the bonding and the nonbonding interactions is somewhat arbitrary: for example, OPLS-AA excludes all nonbonding interactions separated by less than three bonds, and introduces a 0.5 scaling factor for all the 1-4 interaction.[1] Polarizable force field such as AMOEBA adopts an even more complicated exclusion rule based on both topological distances and predefined polarization groups.[11] While how to exclude nonbonding interactions between bonded atoms is subject to personal tastes, an unreasonable exclusion rule may lead to a PES that is too steep or too coupled to fit using simple intramolecular terms. In this work, we will show that machine learning (ML) techniques can be used to solve this problem, due to its extraordinary fitting capability.

Recently, ML emerges as an extremely powerful tool to fit high-dimensional PES. A fairly common practice is to encode the local environment of each atom into a feature vector, and use techniques like artificial neural network (ANN) or kernel-based approaches to predict the corresponding atomic energies.[11–16] The total energy is then expressed as a sum of these local atomic energies. These methods have been extensively used to study inorganic materials and small molecular systems, but their applications to large polymers are relatively rare. Among the existing works, Gradient Domain Machine Learning (GDML) and its symmetrized variant (i.e., sGDML) have been used to develop potentials for medium-sized flexible molecules up to 25 atoms.[17–20] By exploiting symmetry, a CC level force field can be constructed using only a few hundreds of ab initio data points! However, in their work, the features of the entire molecule are fed directly into the ML model to learn the total molecular energy. Therefore, it is unclear how this method can be extended to large polymers, when CC calculation for the entire molecule becomes infeasible. Similar issue exists in the work done by Cole et al.,[21] and Fredrich et al.,[10] in which either Gaussian Approximation Process (GAP) or ANN were employed to develop potentials for molecules no larger than a few tens of atoms. Zhang et al. developed a multi-head attention (MHA) model, named molecular CT,[22] with dual graph and real-space representations. The descriptors and the MHA model used in molecular CT are localized, so what is learned in small systems can be transferred to large systems, in principle. However, no evidence is presented so far that one can build a polymer potential by learning small molecule ab initio data using molecular CT. Similarly, the ANI-1ccx model[23] trains the localized Behler-Parrinello neural network (BPNN)[12] on a vast number of medium-sized molecules at CCSD(T) level of theory, but its transferability to large polymers needs more rigorous proof. Fundamentally speaking, the extendibility of the ML models relies on localized descriptors, which, however, cannot describe long-range nonbonding interactions.

Compare to these models, a probably more extendible approach is to adopt a range-separation scheme, and learn the short-range part using ML. Along this line, Yao et al. developed the TensorMol model, in which BPNN is used, in combination with conventional Coulomb and LJ terms, to fit the PES.[24] Their model has shown excellent transferability to molecules that were not included in the training set, which is a natural reward if the ML model is range-separated and well localized. The remaining issue, though, is that the entire model is built upon DFT, and the fitting quality is also relatively low (as we will compare it with our model below). Its accuracy is mainly limited by the crude treatment of the dispersion and electrostatic interactions, and its transferability with respect to molecular size is yet to be proved. Another work worth noting is done by Cheng et al.,[25] in which fragmented DFT is taken as reference for ML model, in combination with long range Coulomb terms. This approach also takes the advantage of range-separation, and is transferrable to polymer molecules with different sizes. However, the sizes of the fragments are still formidably large for high-level CW calculations, since all non-electrostatic nonbonding interactions must be explicitly computed quantum mechanically within each fragment. Here, we emphasize that the utilization of CW methods is a game changer: when ab initio data can only be collected for systems with no more than a few tens of atoms, how to extrapolation from small molecules to large molecules becomes crucial. Therefore, while the community has made tremendous efforts, a data-efficient and extendible methodology is still in need for the development of polymer force fields with CW accuracy.

In this work, following the idea of range separation, we will combine our accurate physics-driven nonbonding terms with a graph-neural-network-based (GNN) scheme to develop an accurate potential for polyethylene glycol (PEG) polymer. We will show that the conventional physics-driven force field is still extremely important in the era of machine learning, as it can significantly increase the locality and reduce the dimensionality of the learning target. Such localization warrants the high extendibility of the resulting potential, which is critical for the large flexible molecule force field development.

## II. Methodology

*System Definition*

In this work, we use the methyl-capped PEG molecule as a proof-of-concept case to demonstrate our force field development methodology. PEG is selected due to its structural simplicity, while still keeping all the relevant intermolecular physics (i.e., electrostatics, polarization, and dispersion etc.). In practice, PEG with different lengths comprise a series of chemicals with broad industrial applications, ranging from organic solvents to color preservers for terra-cotta warriors.[26–28] A number of conventional force fields were constructed for PEG,[29] showing its broad interests in chemical engineering and materials science. For convenience, we will denote the PEG chains as PEG[n], with n indicating the number of oxygen atoms, as well as the number of repeating "C-O-C" units. Note that people often label PEG molecules using its average molecular weight, so PEG[8] in this study is approximately the PEG-400 in industry. In Figure 1 we illustrate all the PEG chains used in this work: PEG[2] is used to develop intermolecular terms, while PEG[4] monomer is used to develop the intramolecular part. The extendibility of the model is then tested using both PEG[4] and PEG[8].

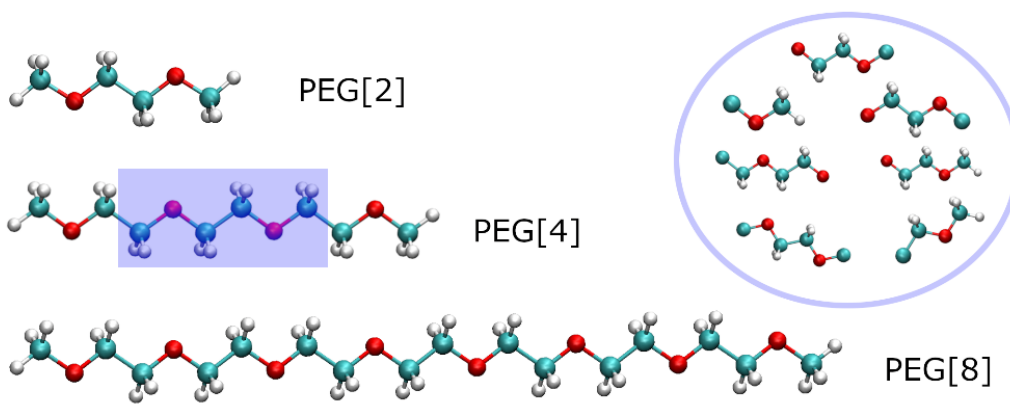

Figure 1. The PEG molecules used in this work: PEG[2] and PEG[4] are used to develop the force fields, while PEG[8] is used to benchmark the accuracy and the transferability of the resulting potential. The blue shaded area labels the approximate size of a subgraph, and within the blue circle are all the subgraphs generated for PEG systems.

*Nonbonding Potential*

For the nonbonding part of the force field, we follow the recipe developed in our previous work.[4–6] We will summarize the procedure very briefly below, and recommend reference 30 and 31 for more mathematical and implementation details. The nonbonding terms are developed via the following two steps:

1. Run TD-DFT calculations on PEG[2] monomer to obtain molecular charge density susceptibility matrix. Then, iterative Stockholder analysis (ISA)[32] and Casimir-Polder relationship are utilized to obtain the distributed atomic parameters including multipoles (up to quadrupole), dipole polarizabilities, and dispersion coefficients ($C_6$, $C_8$, and $C_{10}$).[32] Note that these parameters are in principle conformation-dependent. So we first run a 300 K MD simulation of PEG[2] using OPLS-AA force field, and cluster all sampled structures according to their dihedral distributions into 48 representative conformations. Final atomic parameters are obtained via averaging over the parameters of all 48 conformations. This procedure leads to a potential that matches the SAPT calculations in an excellent accuracy at asymptotic region (see Figure S2). The long-range part of the potential then can be written as:

$$E_{lr} = E_{elec} + E_{ind} + E_{disp} \tag{1}$$

These terms are damped in short-range using Tang-Toennies damping functions:[33,34]

$$E_{elec} = \sum_{i<j} f_1(x_{ij}) \frac{q_i q_j}{r_{ij}} + \sum_{i<j} \sum_{tu} Q_t^i T_{tu} Q_u^j \tag{2}$$

$$E_{disp} = -\sum_{i<j} \sum_{n=3}^{5} f_{2n}(x_{ij}) \frac{C_{ij}^{2n}}{r_{ij}^{2n}} \tag{3}$$

$$f_n(x) = 1 - e^{-x} \sum_{i=0}^{n} \frac{x^k}{k!} \tag{4}$$

$$x_{ij} = B_{ij} r_{ij} - \frac{2 B_{ij}^2 r_{ij} + 3 B_{ij}}{B_{ij}^2 r_{ij}^2 + 3 B_{ij} r_{ij} + 3} r_{ij} \tag{5}$$

Here, the damping exponents $B_{ij}$ are the same as the short-range Slater exponents (*vide infra*), and $Q_t^i$ are the *t*-multipole on site *i*, with $T_{tu}$ being the multipole interaction operators. The induction $E_{ind}$ is described by Drude oscillator model damped by Thole type function,[35] the same as how it is implemented in AMOEBA.[11] Both the pairwise dispersion coefficients $C_{ij}$ and the exponents $B_{ij}$ are computed by taking geometric means of atomic parameters.

2. PEG[2] dimer interaction energies are computed using density fitting DFT-SAPT,[8] Then the medium and short range interactions are fit, using pairwise additive Slater-type functions.[30] The short-range part of the potential is:

$$E_{sr} = \sum_{i<j} A_{ij} P(B_{ij}, r_{ij}) \exp(-B_{ij} r_{ij}) \quad (6)$$

$$P(B_{ij}, r_{ij}) = \frac{1}{3}(B_{ij} r_{ij})^2 + B_{ij} r_{ij} + 1 \quad (7)$$

$$B_{ij} = \sqrt{B_i B_j} \quad (8)$$

In this work, we fit the total short-range interactions, instead of term-by-term as it is done in reference 30. To avoid unphysical negative exchange, we also apply positive constraints on all $A_{ij}$ during the fitting. The atomic exponents $B_i$ are first obtained from the decay rates of the electron densities, then rescaled to achieve better fitting performance. Overall speaking, compare to reference 30, we sacrifice a little generality to gain better fitting quality.

One remaining issue that is potentially important for large molecule force field is the exclusion rule, especially for polarization interactions. Compare to the complicated exclusion rule employed in AMOEBA, our exclusion rule is purely based on connection topological distance. Theoretically, the TD-DFT response matrix already accounts for all intramolecular polarization in the PEG[2] molecule, so the mutual induction within the molecule should be turned off in the force field to avoid double counting.[4] Therefore, considering the size of the PEG[2] molecule, all intramolecular nonbonding interactions within five bonds (*i.e.*, the 1-6 interactions) are excluded, when the PEG[2] parameters are transferred to larger molecules. The exclusion rule may affect the locality of the residual intramolecular energies, so the exclusion distance should match the size of the subgraphs employed in the following GNN model (*vide infra*).

*Subgraph Neural Network (sGNN) Model for Bonding Potential*

Once we have a working nonbonding force field in hand, the residual bonding energy, which is a coupled function of all internal coordinates, is tackled using a message-passing GNN model. MP2/AVTZ calculations on PEG[4] are performed to provide all the training data. We use MP2 in this work as MP2 is feasible for both training and large molecule testing, whilst it also provides a good accuracy compare to SAPT for the PEG molecule (see Figure S1). It is noted that MP2 may fail in describing some systems such as strong dispersions between aromatic rings, so methods like CCSD(T) are often needed.[36] However, the training molecule (*i.e.*, PEG[4]) in the present study contains only around 10 heavy atoms, certainly within the reach of CCSD(T). Therefore, in principle, the procedure can be done at a higher level of theory when necessary.

Once the nonbonding terms are removed from the MP2 energy, we assume the remaining bonding energy can be written as a sum over different local fragments of the molecule. These fragments are defined as "subgraphs" (labeled using letter *g*):

$$E_{sGNN} = \sum_{g} E_g \qquad (9)$$

Each subgraph defines the local environment of a central bond, and $E_g$ represents the bonding energy (including the conventional angle and dihedral terms) attributed to that bond. This leads to a rigorously localized representation of the molecule, warranting the extendibility of the resulting model. Due to such localization, the $E_g$ we learn from PEG[4] subgraphs can be easily transferred to larger polymers containing the same molecular fragments. Different to previous dedicated force fields, such decomposition is similar to BPNN,[12] EANN,[14] and DeepPotential[37] *etc*, which decompose the total energy as a sum of atomic energies. Only in this work, for intramolecular potential, we use internal coordinates, instead of cartesian coordinates as our input features (*vide infra*). Such change of representation may increase the efficiency of the model, as previous work indicates that interatomic distances can be insensitive to certain torsion motions within the molecule.[20] Correspondingly, we also use topological distances, instead of cartesian distances to determine the size of the subgraph: only the first and the second nearest neighboring bonds of the central bond are included in this work. That essentially incorporates all 1-6 interactions, being consistent with the nonbonding exclusion threshold we introduce above. For more coupled molecules, one can further increase the nonlocality of the model by systematically increasing the size of the subgraph. Since the number of subgraphs always equal to the number of bonds, the computational cost of the model always scales linearly with system size, with the prefactor determined by the topological cutoff. In Figure 1, we show all the subgraphs constructed from PEG[4], which can be reassembled to predict the energy of PEG[8] or even longer chains.

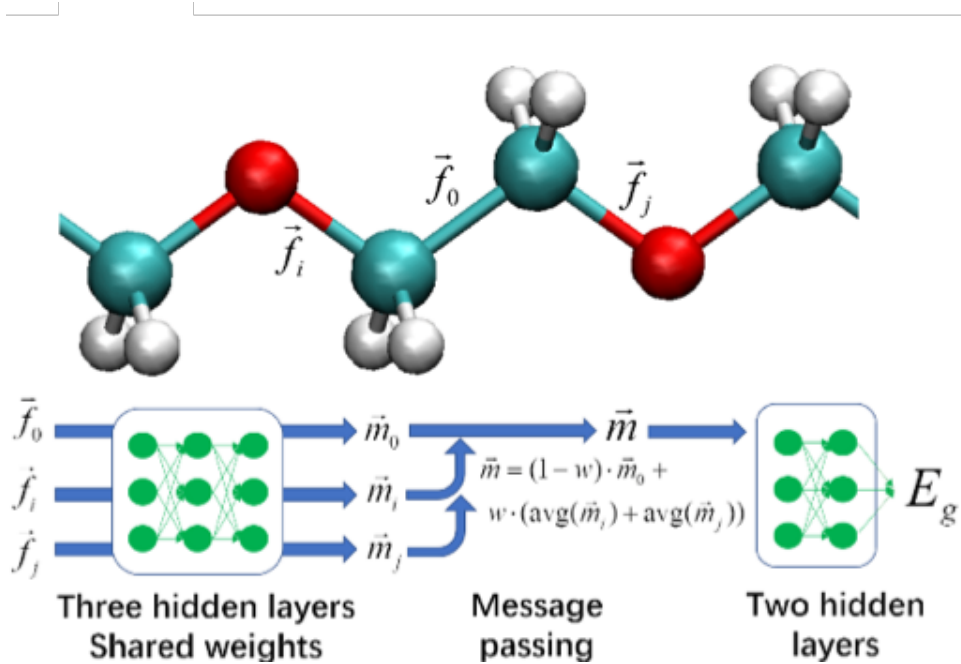

Figure 2. Illustration of the sGNN model, the structure of the message passing GNN.

With subgraphs built, their energies $E_g$ are predicted using a message passing GNN model.[38] The central bond and its nearest neighboring bonds are considered as nodes in the graph, which are connected if they share atom. Each bond is firstly assigned a state vector $f$, which includes the information of the atom types, bond lengths, angles, and dihedrals cosines around this bond. In this way, the three-dimensional structural information of the molecule can be encoded into the state vectors, which can be conveniently fed into ML models.

Then, we adopt a GNN infrastructure to predict energies and forces, including both message passing and aggregation steps. Each state vector first goes through a fully connected several-layer network (here we use three hidden layers) with shared parameters, obtaining the message vector $m$. Then the interactions between the neighboring bonds and the central bond is incorporated by passing $m$ to the central bond via a local weighted average. And a fully connected network with two hidden layers is used to aggregate the message and extract the subgraph energy. This approach can be easily generalized to larger subgraphs: we can do a few more iterations of message passing and aggregation steps, so the central bond senses the information from further bonds. The number of iterations, in conjunction with subgraph sizes, controls the nonlocality of the final model and can be tuned in a systematical way. The whole infrastructure of the sGNN model is illustrated in Figure 2.

*Permutation Symmetry*

Different to physics-driven force fields, for ML models, care must be taken to ensure that all the important physical symmetries are respected. The sGNN model uses internal coordinates as input, so the rotation and translation symmetries are guaranteed. Meanwhile, we further implement the permutation symmetry by taking average of all permuted inputs:

$$E_{sGNN} = \sum_g \frac{1}{P_g} \sum_{p=1}^{P_g} E_g^p \tag{10}$$

Here, $P_g$ stands for all equivalent atom index permutations within subgraph $g$. More specifically, four steps are taken to find all permutations: 1. For each subgraph, a string label is assigned to each atom according to its chemical environment in the subgraph, using the Multilevel Neighborhoods of Atoms (MNA) algorithm;[39] 2. An extra letter is added to the label of hydrogen atoms, to distinguish hydrogens in different chiral positions; 3. All atoms are sorted according to their string labels, and atoms with the same label (*i.e.*, the topologically symmetric atoms) form a permutation group; 4. Atom indices are fully permuted within each permutation group. We note that this permutation algorithm also stands as the key difference between sGNN and other global GNN approaches that treat the entire molecule as a big graph. For a large molecule, a full permutation of all atoms is infeasible, as the computational cost grows as $O(N!)$. However, here we only permute within each permutation group in each small subgraph, so the number of permutations is well under control. On average, the permutation number is less than 6 for all subgraphs, creating minor computational cost increase, due to the strong capability of GPU on vectorized operations.

Eventually, the energy and the force for the entire molecule can be computed by combining the long-range nonbonding, the short-range nonbonding, and the sGNN bonding terms:

$$E = E_{lr} + E_{sr} + E_{sGNN} \tag{11}$$

### III. Computational Details

All SAPT and MP2 calculations were performed using Molpro 2019 program,[40,41] with Dunning style AVTZ (aug-cc-pVTZ) basis set.[42–44] For dimer interaction calculations, even-tempered basis functions (5s5p3d2f) are placed in the midpoint between the centers of mass, with exponents ratio set to be 2.5, and centering at 0.5, 0.5, 0.3, 0.3 a.u. for the s, p, d, f shells, respectively. LPBE0AC functional was used for the df-DFT-SAPT calculations.[8] Asymptotic nonbonding parameters were obtained using the CAMCASP 6.1 program, with ISA-pol population analysis method.[32,45] PBE0 density functional in conjunction with ALDA kernel was used, interfacing with NWChem[46] for the TD-DFT response calculations. To develop short-range nonbonding interactions, we ran 50 different PEG[2] dimer scans, with dimer conformations sampled using OPLS-AA MD simulation.

To train sGNN, we ran 20,000 MP2 calculations on PEG[4] in total. The geometries were sampled from 50 ns OPLS-AA MD simulations, conducted at both 300 and 1000 K. 10% of the entire dataset were drawn as testing set, and 90% of it were kept as training data. Testing conformations for PEG[8] were also generated via a 50 ns OPLS-AA MD simulation at 300 K. For the training process, we employed ADAM optimizer,[47] with minibatch size set to either 16 (when we only train energies) or 8 (when force data is included in training). Unless stated otherwise, the learning rate was set to 0.0001, and we ran the optimizations for 4000 epochs at maximum. Features were centered and scaled to make the distribution roughly within (-1, 1), and the target energy is also shifted and scaled to standard normal distribution.

The nonbonding energies were computed using the AMOEBA plugin in OpenMM 7.4,[48] with small modifications to the source code to correctly exclude all 1-6 interactions. The bonding terms were implemented using PyTorch, and two i-Pi drivers (interfacing with OpenMM and Pytorch, respectively)[49] were developed to enable us to run MD simulations using i-Pi.[50] The sGNN MD simulations were run in NVT ensemble, utilizing Langevin thermostat with a timestep of 0.5 fs.

## IV. Results and Discussions

*Nonbonding Potential*

Due to the physics-driven nature of the bonding interactions, the nonbonding part of the force field can be developed with a very low computational cost. In total, only 48 TD-DFT calculations and 600 PEG[2] dimer SAPT calculations are needed. The final parameters in use are given in the supporting materials, and the fitting results are shown in Figure S2. The total root mean squared error (RMSE) is 0.23 kcal/mol, similar to what has been reported in reference 30. Most error comes from the short-range interactions, as the RMSE is 0.31 kcal/mol for the short-range geometries (geometries with the closest contact shorter than 3.2 Å), and 0.06 kcal/mol for others. While the exact long-range behavior guarantees the transferability, the short-range interactions sets the upper limit to the accuracy of the model. Therefore, more work is needed to refine the short-range nonbonding interactions in future. Nevertheless, as we will show, the accuracy for the nonbonding potential achieved in this work is already high enough for a robust PEG force field.

*Bonding Potential Training Results*

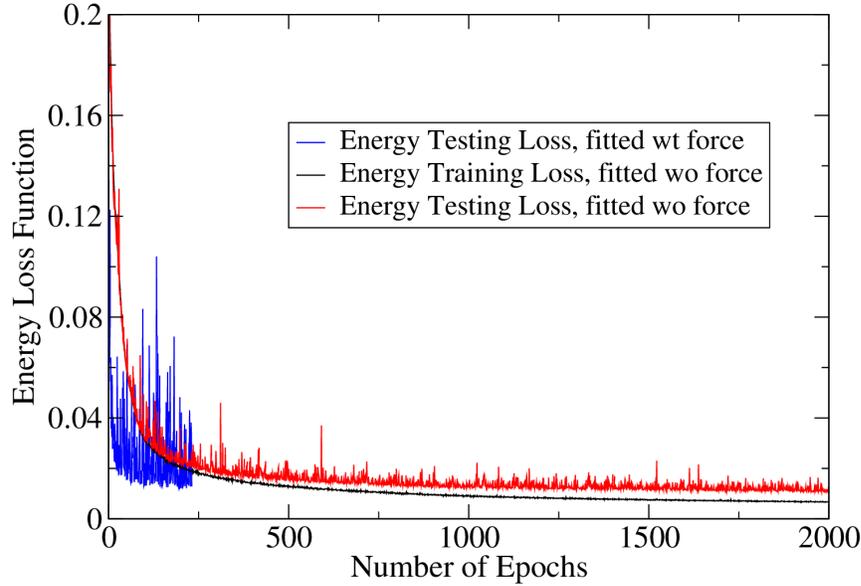

Figure 3. The training and testing energy loss functions during a training process. Both training and testing loss functions are computed on the PEG[4] system.

The trends of the training and testing energy loss functions in one typical training process is shown in Figure 3. The energy loss function is simply defined as:

$$L_{energy} = \frac{1}{N} \sum_i \left(\varepsilon - \varepsilon_{ref}\right)^2 \qquad (12)$$

In which $N$ stands for the number of data points, and $\varepsilon$ stands for normalized energies. We can see that both the training and testing losses keep decreasing while training, without signs of significant overfitting. As shown in Figure 3, adding forces to the loss function can significantly accelerate the training process. Facilitated by force matching, the testing energy loss quickly converges within a few hundreds of epochs, with some fluctuations caused by the constant shifts of the PES. The final fitting qualities are comparable for both force matching and energy matching schemes. However, we note that in this case, the major computational bottleneck is not the training process, but the training data collection. For nonvariational CW methods like MP2 or CC, the analytical force calculations can be much slower than simple energy evaluation. Consequently, different to previous DFT-based work, we find it slightly more efficient to only fit to energies, which is what we are going to do in the remaining part of the paper. However, the force matching capability is still implemented in the code as it could be important for accurate dynamics or spectroscopy calculations.

*PEG[4] and PEG[8] Tests*

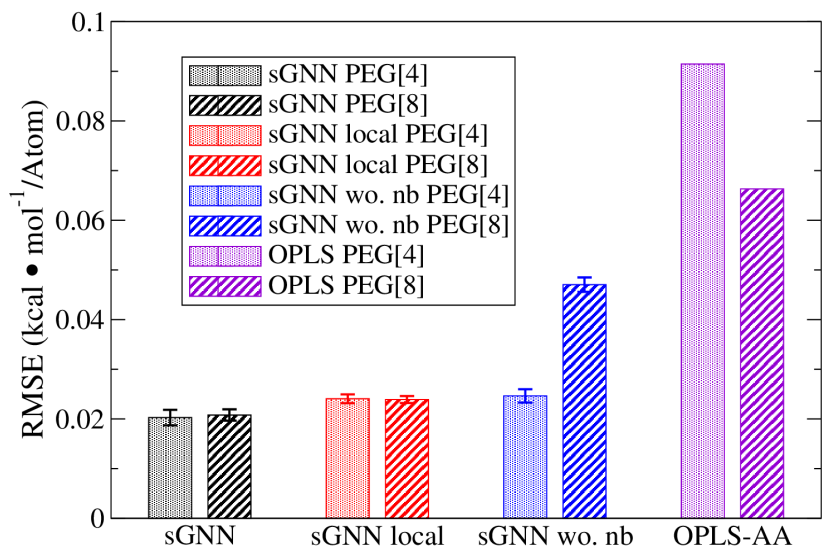

Figure 4. Summary of the testing RMSEs of different sGNN models, in comparison with OPLS-AA. Error bars indicate the standard deviation among different fittings.

Once we train the models on PEG[2] and PEG[4], we test their accuracies on both PEG[4] and PEG[8], by checking their energy RMSEs comparing with MP2. For each type of model, we always conduct five independent fittings, and the average RMSE is reported with the standard deviation among the five fittings. The RMSE results are shown in Figure 4, and the direct energy comparisons are shown in Figures S3.

For both PEG[4] and PEG[8], the sGNN RMSE is around 0.020 kcal•mol$^{-1}$/atom, much smaller than that of OPLS-AA and TensorMol-1.0 (which features a typical RMSE of 0.054-0.24 kcal•mol$^{-1}$/atom).[24] The small RMSE on PEG[8] proves that the sGNN model is highly transferrable to large molecules, even though it was trained on small ones. This success is nontrivial, since in spite of their similar chemical structures, PEG[4] and PEG[8] are very different molecules. To illustrate this, we plot the end-to-end distance distributions of PEG[4] and PEG[8] in Figure S4. Even though the PEG[8] chain is longer, its most probable end-to-end distance is much shorter than that of PEG[4]. This is because due to the rigidity of the polymer skeleton, PEG[4] is too short to form any folded structures, while PEG[8] can fold. Folded structures feature nonbonding contacts between different parts of the chain, which is captured by the nonbonding part of our force field.

Here, we emphasize that the separation between the nonbonding and the bonding terms is critical, as the nonbonding part cannot be learned from the small PEG[4] molecule. To further illustrate this point, we fit our sGNN model without removing the nonbonding interactions, and show the testing results in Figure 4 (labeled as "sGNN wo. nb"). While such model performs reasonably well for PEG[4] (with a RMSE of 0.025 kcal•mol$^{-1}$/atom), it deteriorates rapidly in the PEG[8] test (with a RMSE of 0.047 kcal•mol$^{-1}$/atom). Interestingly, the RMSE for all the extended conformations of PEG[8] (conformations with an end-to-end distance larger than 13 Å) is much smaller (0.021 kcal•mol$^{-1}$/atom). This proves that the transferability problem is indeed caused by the nonbonding contacts in the folded structures, illustrating the importance of an accurate nonbonding description. One can of course try to train the potential using a larger molecule, with all the short-range nonbonding interactions included. But such strategy is very inefficient, especially considering the poor scaling of the CW methods. Therefore, we strongly advocate the methodology that hybrids physics-driven potentials and ML potentials, so the advantages of both methods can be exploited.

As discussed in the Methodology section, the couplings between internal coordinates can be systematically tuned in sGNN by changing the subgraph size and the number of message passing/aggregation iterations. Here, we examine how important such nonlocal coupling is, by skipping the message passing step (resulting a model labeled as "sGNN-local" in Figure 4). It is noted that even without the message passing step, the state vector $f$ for each bond still encodes the information of all its neighboring atoms, thus accounting for all 1-4 interactions. But any couplings beyond three covalent bonds would be missing. The PEG[8] RMSE for the sGNN-local model is $0.024 \pm 0.0007$ kcal•mol$^{-1}$/atom, in comparison with the $0.021 \pm 0.001$ kcal•mol$^{-1}$/atom for the full sGNN model. The difference is small but noticeable, demonstrating that while PEG is a simple molecule, there still exists some level of nonlocal couplings. Such nonlocal coupling can be important for other molecules with highly correlated bonding interactions such as intramolecular hydrogen bonds. Our sGNN model thus provides an infrastructure with an easy handle to capture and tune such nonlocality in the bonding force field.

Since our sGNN model is trained by matching the energy of the entire PEG[4] molecule, one important issue is whether sGNN faithfully reproduces the local PES of each subgraph, or it is relying on the error cancelations between different subgraphs. Such "local fidelity" is crucial if one wants to transfer the subgraph to a different molecule with the same local chemical structure. Therefore, besides the MD sampled conformations, we also run a dihedral scan over the central bond of PEG[4], and the results are shown in Figure 5. It can be seen that sGNN predicts a smooth curve that accurately captures both the high and the low dihedral rotation barriers. It means that while being trained in one sum, each subgraph captures the corresponding local conformation energy accurately. Therefore, it can be taken out as a single term and used in a different molecule.

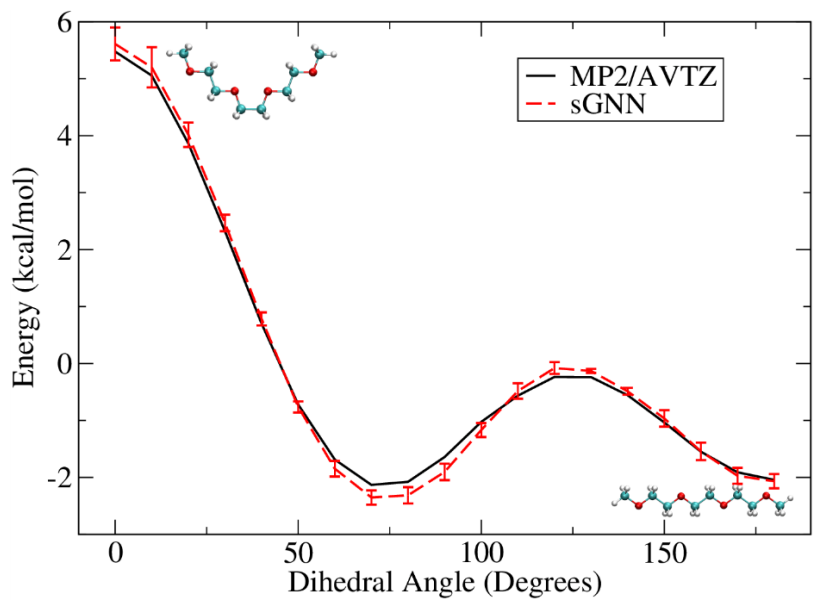

Figure 5. Potential energy scan on the central dihedral of PEG[4], sGNN versus MP2 calculation results. Error bars indicate the standard deviation among different fittings.

*MD Simulation Tests*

To examine the stability of the sGNN model in real MD, we run a 100 ps NVT simulation of a single PEG[8] chain, starting from the fully extended geometry. Five models from five independent fittings are used in each MD step to predict the energy and the force, while the dynamics is propagated on the averaged PES. To check the fidelity of the potential on the fly, the self-consistency among the five models is computed on each step, and a warning would be raised if their relative difference is larger than 10%.

As shown in Figure 6, the potential energy stays stable throughout the 100 ps simulation, and the conserved energy reported by i-Pi fluctuates around 0.044 a.u. without drifting, proving the smoothness of the PES. Moreover, the energy inconsistency among different models stays at a constant level of about 5% (*i.e.*, 0.03 kcal/mol). Therefore, even though the PEG[8] molecule samples a very diverse set of folded and unfolded structures, they are all well covered by our training dataset. We find the instability issue commonly seen in other ML infrastructures rarely happens in sGNN. The subgraphs contain no more than 12 atoms, so the dimensionality of sGNN is much lower than previous works, leading to a model that is much easier to train. This is again rooted to the careful separation of the nonbonding interactions from the ML model, which leads to an effective localization of the problem and brings a significant improvement to the numerical performance.

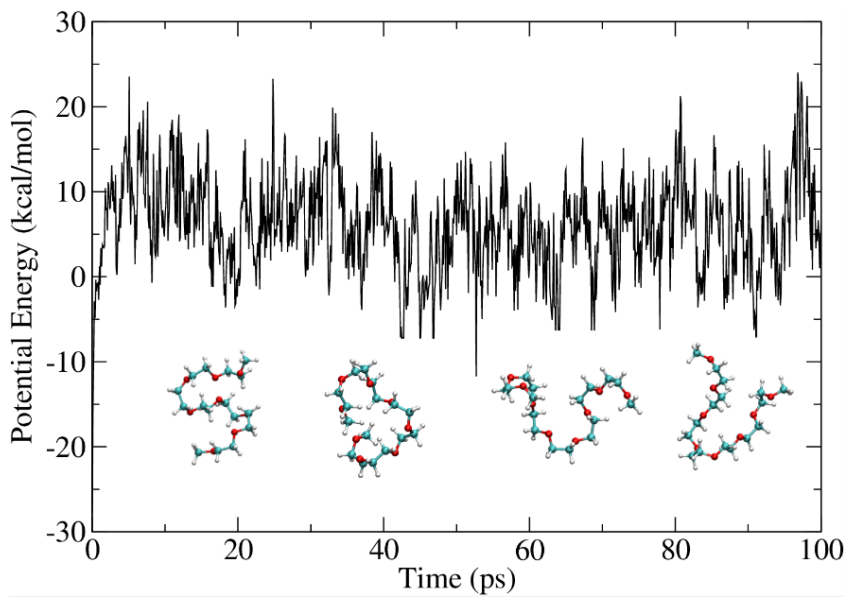

Figure 6. Potential energy trajectory of the sGGN NVT simulation, and the conformations sampled by MD.

## V. Outlooks and Conclusions

In this work, we developed a GNN-based strategy (named sGNN) to describe the PES of large flexible polymer molecules. The new recipe contains three key ingredients: a clean separation of bonding and nonbonding interactions; a decomposition of large molecules into small subgraphs;

and a powerful message passing GNN model. Using PEG as example, we demonstrate that the newly developed sGNN model is not only accurate, but more importantly, extendible to molecules with larger sizes. It is further shown that such extendibility relies on the accurate nonbonding potential obtained from TD-DFT and SAPT calculations, so the physics-driven nonbonding terms play a key role in the success of sGNN. Essentially, we show that after carefully removing the nonbonding interactions, the residual bonding energies are highly localized, thus can be well described using very small subgraphs. Therefore, by combining the physics-driven nonbonding potential with the data-driven bonding potential, we exploit the advantages of both methods. Such advantages allow us to construct large molecule force fields from very small high-level ab initio calculations, thus circumventing the scaling curse in quantum chemistry.

As a proof of concept, only one type of polymer molecule (*i.e.*, PEG) is examined in this work, but there is no fundamental obstacle to generalize sGNN to a larger chemical library. A unique advantage of sGNN is that the subgraph PES is most likely to be transferrable to a different molecule with the same fragment. Therefore, we can potentially develop the force field using an incremental strategy: when new molecules are introduced, we only need to train the new subgraphs, while keeping the potential of the previously seen subgraphs unchanged. When training new subgraphs, they do not even need to share parameters with old subgraphs, as they can be trained separately. In this way, the training cost increases linearly with the number of subgraph types, instead of molecule types, while the number of molecule types increases exponentially with the number of subgraph types. Therefore, sGNN can be a very robust and cheap framework for developing general purpose organic force fields beyond DFT accuracy.

ASSOCIATED CONTENT

**Supporting Information**.

The following files are available free of charge.

Supplementary figures and short-range nonbonding parameters. (PDF)

Asymptotic atomic parameters, including atomic multipoles, dispersion coefficients, polarizabilities, provided in Openmm force field input format. (XML)

AUTHOR INFORMATION

**Corresponding Author**

*Kuang Yu - Tsinghua-Berkeley Shenzhen Institute (TBSI), Institute of Materials Research (iMR), Tsinghua Shenzhen International Graduate School (TSIGS), Tsinghua University. 1305 Information Building, University Town, Shenzhen, Guangdong Province, China, 518055; Email: yu.kuang@sz.tsinghua.edu.cn

**Author Contributions**

The manuscript was written through contributions of all authors. All authors have given approval to the final version of the manuscript.

**Notes**

The authors declare no competing financial interest.

ACKNOWLEDGMENT

The authors acknowledge Professor Xiaolong Zou for useful discussions, and Powerleader Science & Technology Group for the technical supports in the maintaining our high-performance computers.

ABBREVIATIONS

DFT: density functional theory;

CW: correlated wavefunction methods;

MP2: Second order Moller-Plesset perturbation;

CCSD(T): coupled cluster with single, double, and perturbative triple excitations;

SAPT: symmetry-adapted perturbation theory

ML: machine learning

sGNN: subgraph neural network model

MD: molecular dynamics

MC: Monte Carlo

PEG: polyethylene glycol